\documentclass[12pt]{article}
\usepackage{cite,epsfig,amssymb,amsmath,graphicx,subfigure,color}
\usepackage{caption}
\evensidemargin \oddsidemargin
\evensidemargin \oddsidemargin

\newcommand{\nn}{\nonumber}

\begin{document}

\begin{center}
{
{{\Large \bf  Renormalized Holographic Subregion Complexity under Relevant Perturbations}
}\\[17mm]
Dongmin Jang$^{1}$,~~Yoonbai Kim$^{1}$,~~O-Kab Kwon$^{1}$
~~D.~D. Tolla$^{1,2}$\\[3mm]
{\it $^{1}$Department of Physics,~BK21 Physics Research Division,
 Autonomous Institute of Natural Science,~Institute of Basic Science, Sungkyunkwan University, Suwon 440-746, South Korea\\
$^{2}$University College,
Sungkyunkwan University, Suwon 440-746, South Korea}\\[2mm]
{\it dongmin@skku.edu,~yoonbai@skku.edu,~okab@skku.edu,~ddtolla@skku.edu} }

\end{center}
\vspace{15mm}

\begin{abstract}
We construct renormalized holographic entanglement entropy (HEE) and subregion complexity (HSC) in the CV conjecture for asymptotically AdS$_4$ and AdS$_5$ geometries under relevant perturbations. Using the holographic renormalization method developed in the gauge/gravity duality, we obtain counter terms which are invariant under coordinate choices. We explicitly define different forms of renormalized HEE and HSC, according to conformal dimensions of relevant operators in the  $d=3$ and $d=4$ dual field theories.  We use a general embedding for arbitrary entangling subregions and showed that any  choice of the coordinate system  gives the same form of the counter terms, since they are written in terms of curvature invariants and scalar fields on the boundaries. We show an explicit example of our general procedure. Intriguingly, we find that a divergent term of the HSC in the asymptotically AdS$_5$ geometry under relevant perturbations with operators of conformal dimensions in the range $0< \Delta < \frac{1}{2}\,\, {\rm and} \,\, \frac{7}{2}< \Delta < 4$ cannot be cancelled out by adding any coordinate invariant counter term. This implies that the HSCs in these ranges of the conformal dimensions are not renormalizable covariantly.

\end{abstract}

\newpage
\tableofcontents

\section{Introduction}
The holographic realizations of the entanglement entropy and the quantum complexity have established a  connection between  the gravity theory and  quantum information theory through gauge/gravity duality~\cite{Maldacena:1997re, Gubser:1998bc, Witten:1998qj}. The Ryu-Takanayagi (RT) conjecture~\cite{Ryu:2006bv, Hubeny:2007xt} laid out a holographic way of calculating the entanglement entropy for a subregion $A$ in a $d$-dimensional field theory on the boundary of a $(d+1)$-dimensional bulk geometry of the dual gravity theory. According to the RT conjecture, the holographic entanglement entropy (HEE) is proportional to the area of $(d-1)$-dimensional bulk minimal hyper-surface $\Sigma_A$, which is homologous to the subspace $A$ in a $d$-dimensional constant time slice. The quantum complexity is also an important quantity in the information theory, which measures how many minimum simple gates to reach from a simple reference state to a target state. However, the notion is not well-defined in quantum field theory generally. There have been two proposals to calculate the quantum complexity in terms of the gauge/gravity,  which are referred to as the CV (Complexity=Volume) 
conjecture~\cite{Stanford:2014jda} and the CA (Complexity=Action) 
conjecture~\cite{Brown:2015lvg,Brown:2015bva}. These correspond to the complexity of a pure state in the whole boundary space of the dual quantum field theory. Natural generalizations of the CV and CA conjectures are holographic complexities of a mixed state for the reduced density matrix for a subregion $A$. These are known as the holographic subregion complexity (HSC) for the CV conjecture~\cite{Alishahiha:2015rta} or the CA conjecture~\cite{Carmi:2016wjl}. Other studies on the HSC include 
\cite{Alishahiha:2015rta, Ben-Ami:2016qex, Carmi:2016wjl, Roy:2017kha, Bakhshaei:2017qud, Roy:2017uar, Abt:2017pmf, Chen:2018mcc, Agon:2018zso, Abt:2018ywl, Alishahiha:2018lfv, Karar:2019wjb,  Bhattacharya:2019zkb, Auzzi:2019fnp,Ghosh:2019jgd, Auzzi:2019mah, Caceres:2019pgf, Geng:2019yxo}. In this paper, we construct renormalized HEE and HSC by focusing on the CV conjecture and hence the HSC in this paper refers to the quantity obtained through the CV conjecture. 

The HSC states that the quantum complexity of a mixed state, which is produced by reducing the boundary state to a specific subregion $A$, is proportional to the volume of the extremal hyper-surface ${\cal B}_A$ enclosed by the boundary subregion $A$ and corresponding RT surface $\Sigma_A$. See Fig.1. Therefore, in order to calculate the HSC, one has to fix the RT surface at first by solving equations of motion to minimize the codimension two hyper-surface ending on the boundary of the subregion $A$. Then the HSC is computed by 
\begin{align}\label{CA0}
{\cal C}_A = \frac{V({\cal B}_A)}{8 \pi L G_{d+1}},
\end{align}   
where $L$ and $ G_{d+1}$ are the AdS radius and the Newtonian constant in $(d+1)$-dimensions, respectively. It was also suggested that the quantity ${\cal C}_A$ in \eqref{CA0} can be interpreted as the fidelity susceptibility in the quantum information theory~\cite{Alishahiha:2015rta, MIyaji:2015mia}.

The HEE and  HSC involve integration over extremal subspaces that are extending to the asymptotic boundary. As a result, they are divergent due to the infinite area/volume of the extremal subspaces on the boundary.  In the dual boundary field theory, these divergences correspond to the UV divergences, which are related to the short distance correlations, and it is necessary to renormalize those holographic quantities. One well-known method to renormalize the HEE is to cancel out the divergent terms by using differentiation with respect to a characteristic length scale of the entangling 
subregion~\cite{Liu:2012eea}. See also \cite{Nishioka:2018khk}. However, this method depends on the shape of the entangling region and the choice of coordinate system. In order to overcome the disadvantages of the differentiation method, a systematic renormalization method known as the holographic renormalization~\cite{Emparan:1999pm, deHaro:2000vlm, Skenderis:2002wp, Bianchi:2001kw} was applied to the renormalization of the 
HEE~\cite{Taylor:2016aoi}.  See also \cite{Anastasiou:2018rla, Anastasiou:2019ldc}. Application of this method to the HSC for pure AdS spaces was also discussed in \cite{Carmi:2016wjl}. Renormalization of the holographic complexity for pure states in terms of the holographic renormalization method was studied in \cite{Kim:2017lrw}.

In this paper, we construct renormalized HEE and HSC with    arbitrary entangling subregions for asymptotically AdS$_4$ 
and AdS$_5$ 
geometries~\footnote{We omit the case of the  AdS$_3$ geometry, which is  similar to the case of the AdS$_5$ geometry.} under relevant perturbations by introducing a scalar field. We determine covariant counter terms on the cut-off boundary in terms of   the holographic renormalization method.  Our construction also can be applied to generic asymptotically AdS geometries, such as AdS black holes and AdS solitons, etc.

In the case of the HEE, recalling that the RT minimal hyper-surface $\Sigma_A$ is homologous to the subspace $A$, its boundary $\partial\Sigma_A$ is independent of the bulk stress tensor. However, the subleading divergences in the HEE are determined by the back reaction of the stress tensor on the geometry. In order to account for these subleading divergence, the counter terms should contain invariants of the matter fields in addition to the curvature invariants on $\partial\Sigma_A$. We determine the exact forms of these counter terms in  asymptotically AdS$_4$ and AdS$_5$ geometries under relevant perturbations. The renormalized HEE with a disk entangling subregion in  asymptotically AdS$_4$  geometry under relevant perturbations was obtain in \cite{Taylor:2016aoi}. This result is also obtained from our result of the asymptotically AdS$_4$ geometry.

In the case of the HSC, the curvature invariants on the cut-off boundary are dependent on the bulk stress tensor. Therefore, the counter terms to cancel the leading as well as subleading divergences can be the integrals of the curvature invariants on the  $(d-1)$-dimensional cut-off boundary.  However, since the $(d-1)$-dimensional cut-off boundary
meets the $(d-2)$-dimensional  boundary $\partial\Sigma_A$ of the RT hyper-surface, there is always one divergent term, which is expressed in terms of integrals of curvature invariants on $\partial\Sigma_A$. This peculiar divergence is logarithmic when $d$ is odd and is a power-law divergence when $d$ is even. We show that the complete counter terms for the HSC are expressed as integrals of the curvature invariants on the $(d-1)$-dimensional cut-off boundary plus integrals of the curvature invariants on the $(d-2)$-dimensional boundary of the RT hyper-surface. We apply this procedure to a particular example of an asymptotically AdS$_4$ geometry obtained from the non-linear KK reduction of the 11-dimensional LLM geometry~\cite{Lin:2004nb} and obtain coordinate independent finite results.

Intriguingly, we find that there exist a divergent term ${\cal O} (\epsilon^{\alpha-1})$ with the range $0<\alpha<1$, which cannot be cancelled out by adding any curvature invariant, in the renormalization procedure of the HSC in the asymptotically AdS$_5$ geometry under relevant perturbations. This implies that there is no renormalized HSC in the range $0<\alpha<1$, with this range of $\alpha$ corresponding to the conformal dimension of the relevant operators in the 4-dimensional dual field theory to be in the range $0< \Delta < \frac{1}{2}\,\, {\rm and} \,\, \frac{7}{2}< \Delta < 4$. Here we also notice that the latter case does not violate the unitary bound $(\Delta \ge1)$ for primary operators. It will be interesting if one figures out the physical reason of this phenomenon. To do that, one needs more investigations for other HSC, such as in the CA conjecture and other dimensions to resolve this problem.

The remaining parts of this paper are organized as follows. In section 2, we discuss the renormalization of HEE in an asymptomatically AdS$_4$ and AdS$_5$ geometries, which are obtained from the perturbation of the AdS geometries with a scalar field. We show the counter terms are determined by the curvature invariants on the boundary of the RT hyper-surface as well as the scalar field. In section 3, we renormalize the HSC. We point out that the counter terms built just from the curvature invariants on the cut-off boundary are not enough to cancel the  divergences and show the need for including the curvature invariants on the boundary of the RT hyper-surface. We obtain the  forms of the counter terms in the asymptomatically AdS$_4$ and AdS$_5$ geometries. In section 4, we apply the general results of section 2 and 3 to the KK reduction of the LLM geometry. We draw our conclusions in section 5.

\section{ Renormalized HEE   under Relevant  Perturbations} 

The UV divergences
in the EE and the quantum complexity naturally appear due to the strong entanglement near the boundary of entangling regions in quantum field theory. In the dual gravity theory, the corresponding HEE and HSC  also have the corresponding UV divergences, which should be renormalized  before we associate  physical phenomenon with the entanglement and the quantum complexity. In this section, we focus on  the renormalization of the HEE in asymptotically  AdS$_{d+1}$ geometries, which are obtained by relevant perturbations that correspond to insertion of scalar fields in the dual gravity. A simple well-known way to renormalize the HEE is to use the differentiation for the HEE with respect to a characteristic length scale~\cite{Liu:2012eea}, for instance, the radius of the disk or the width of the strip of entangling regions. However, this method cannot be applicable for some entangling regions and depends on the choice of the spacetime coordinate. To overcome these drawbacks, a systematic way was 
proposed~\cite{Taylor:2016aoi} by adopting the method of the holographic 
renormalization~\cite{Emparan:1999pm, deHaro:2000vlm, Skenderis:2002wp, Bianchi:2001kw} in the gauge/gravity duality. For the relevant perturbation near the asymptotic region, the renormalized HEE for a disk in the asymptotically  AdS$_{4}$ geometry was obtained.   
Here we briefly review the 
method~\cite{Taylor:2016aoi}, and extend the method to some entangling regions for the asymptotically AdS$_4$ and AdS$_5$  under relevant perturbations.
    
In the next section, we consider the renoramalization of the HSC with arbitrary shapes of entangling regions on asymptotically AdS$_4$ and AdS$_5$ geometries.  We propose new counter terms, which are genuine in the  renormalization of the HSC.  

\subsection{Renormalized HEE in asymptotically  AdS$_4$ geometry}\label{D4}

In the field theory, the relevant deformation of the $d$-dimensional CFT refers to inserting gauge invariant operators with conformal dimension $\Delta <d$, whereas in the dual gravity, this relevant perturbation is achieved by introducing a scalar field with mass $M^2 =\Delta (\Delta -d)$ with $\Delta <d$. Therefore, for the holographic description of the EE under  relevant perturbations in $d$-dimensional CFT, we start from the $(d+1)$-dimensional gravity action
with negative cosmological constant coupled to a scalar field $\phi$,
\begin{align}\label{Sgf}
S_{g\phi}=\frac1{16\pi G_{d+1}}\int d^{d+1}x\sqrt{-g}\left({\cal R}-2\Lambda -\frac12\partial_p\phi\partial^p\phi -\frac12M^2\phi^2 - V(\phi)\right),
\end{align}
where $G_{d+1}$ is the $(d+1)$-dimensional Newton's constant, $x^p=(z, x^\mu )$  with the $d$-dimensional boundary coordinates $x^\mu$ are the bulk coordinates with the holographic radial direction $z$,  $\Lambda=-\frac{d(d-1)}{2L^2}$ is the cosmological constant, and $V(\phi)$ denotes the potential with higher order self-couplings of the scalar field. Under the assumption of the Poincar\'e invariance for the coordinate $x^\mu$, the metric of an asymptotically AdS$_{d+1}$ geometry in the Fefferman-Graham coordinate system is given by
\begin{align}\label{FG1}
ds^2= g_{pq}dx^p dx^q=\frac{L^2}{z^2}\Big(dz^2+\big(1+h(z)\big) \eta_{\mu\nu}dx^\mu dx^\nu\Big).
\end{align} 
The metric fluctuation $h(z)$, which vanishes at the boundary $(z=0)$, measures the deviation from the pure AdS$_{d+1}$ geometry, due to the nonvanishing contribution of the Poincar\'e invariant scalar field $\phi = \phi (z)$.

Plugging \eqref{FG1} and the scalar field ansatz  $\phi = \phi (z)$ into the Einstein equation  and the equation of motion for the scalar field in 4-dimensions, we obtain 
\begin{align}\label{EOM}
&2z (1+h) h''-2 z h'^2+2 (1+h) h'+z(1+h)^2 \phi'^2 =0, 
\nn \\
& {z^2 (1+h) \phi ''}-{2 z (1+h) \phi '}+\frac32 z^2 h' \phi '-L^2M^2 (h+1) \phi+\cdots=0,
\end{align} 
where  the ellipses  denote contributions from the potential $V(\phi)$. In the asymptotic  region $(z\to 0)$, there are two independent solutions of the equations in \eqref{EOM}, 
\begin{align}
&\phi_a(z)=s_0z^{3-\Delta}+s_{1}z^{3(3-\Delta)}+\cdots\quad \Longrightarrow\quad h_a(z)=-\frac{s_0^2}8z^{2(3-\Delta)}+h_1z^{4(3-\Delta)}+ \cdots, 
\nn \\
&\phi_b(z)=v_0z^{\Delta}+v_{1}z^{3\Delta} + \cdots\quad \quad \quad ~\Longrightarrow\quad h_b(z) =-\frac{v_0^2}8z^{2\Delta}+\tilde h_1z^{4\Delta} +\cdots, \label{gh(z)}
\end{align}
where $\Delta=\frac12\big(3+\sqrt{9+4M^2L^2}\big)$ is the conformal dimension of a gauge invariant operator dual to the scalar field $
\phi$. The operator is relevant for $\Delta<3$. According to the GKP-W relation~\cite{Gubser:1998bc,Witten:1998qj} in the gauge/gravity duality, we have the relations
\begin{align}\label{s0v0}
s_0 \sim J_{{\cal O}^\Delta},\qquad v_0\sim \langle {\cal O}^\Delta\rangle, 
\end{align} 
where $J_{{\cal O}^\Delta}$ and $\langle {\cal O}^\Delta\rangle $ are the source and the vacuum expectation value of a gauge invariant operator ${\cal O}^\Delta$ with conformal dimension $\Delta$, respectively. 
The coefficients  ($s_1, h_1,\cdots$) in \eqref{gh(z)} are determined in terms of $s_0$ and  ($v_1,\tilde h_1,\cdots$) are determined in terms of $v_0$, by inserting these solutions into \eqref{EOM} and solving order by order in  $z$. We will show that, the way these solutions depend on the holographic coordinate $z$,  determines the forms of the divergent terms in the HEE and HSC as well as the forms of the appropriate counter terms.

The RT conjecture states that the HEE of a subregion $A$ and its complement $A^c$, which lies on the boundary of ($d$+1)-dimensional bulk geometry, is given by 
\begin{align}\label{S_A}
S_A = \frac{{\rm Min}({\cal A}_{\Sigma_A})}{4 G_{d+1}}= \frac{1}{4 G_{d+1}}\int d z\int d^{d-2}\sigma^a \sqrt{\gamma},
\end{align}
where ${\rm Min}({\cal A}_{\Sigma_A})$ denotes the minimal area of a bulk static hyper-surface $\Sigma_A$, which has the same boundary with the subregion $A$. The induced metric on  $\Sigma_A$ with the target space metric $g_{pq}$ in  \eqref{FG1} is defined as $\gamma_{\alpha\beta}=\frac{\partial x^p}{\partial\sigma^\alpha}\frac{\partial x^q}{\partial\sigma^\beta} g_{pq}$ with the worldvolume coordinate $\sigma^\alpha = \{z,\sigma^a\}$, $a=1,\cdots, d-2$.  
In \cite{Taylor:2016aoi}, the authors introduced a  parametrization  of the embedding of the static surface $\Sigma_A$ at a constant time $t=t_0$ with arbitrary entangling region into the bulk space by setting 
\begin{align}\label{embed0}
x^a = \sigma^a\quad  {\rm and}\quad  x^{d-1} \equiv y= w(z,x^a).
\end{align}
It seems that the embedding \eqref{embed0} with $(d-1)$-dimensional independent parameters  can express most shapes of entangling subregions, since the resulting hyper-surfaces  $\Sigma_A$'s are also  $(d-1)$-dimensional geometries. In this sense, the embedding \eqref{embed0} is applicable to arbitrary shapes of entangling subregions.

For $d=3,$ this embedding is written as
\begin{align}\label{wofzx}
x^p=\Big(t=t_0,z,x,y=w(z,x)\Big).
\end{align}
The induced metric  on the surface $\Sigma_A$  is then given by
\begin{align}\label{indMet}
\gamma_{zz}=\frac{L^2}{z^2}\Big(1+w'^2(1+h)\Big),\qquad \gamma_{zx}=\frac{L^2}{z^2}{\dot w}w'(1+h),\qquad \gamma_{xx}=\frac{L^2}{z^2}(1+h)\Big(1+{\dot w}^2\Big),
\end{align}
where $\sigma^\alpha=(z,x)$ are coordinates on the surface, and we have used the notations $w'=\partial_z w$, ${\dot w}=\partial_x w$.
Then the area of the surface $\Sigma_A$ is calculated as
\begin{align}\label{Area4}
{\cal A}_{\Sigma_A}=\int dx \int dz\sqrt{\det{\gamma_{\alpha\beta}}}=L^2\int dx \int dz\frac{(1+h)^{1/2}}{z^2}\sqrt{1+{\dot w}^2+(1+h)w'^2}. 
\end{align}
The following Euler-Lagrange equation derived from this action determines the minimal area surface:
\begin{align}\label{minEq}
&z(1+h)w''+z{\ddot w}-2(1+h)w'+\frac32zh'w'\nn\\
-&\frac z{{1+{\dot w}^2+(1+h)w'^2}}\left({\dot w}^2{\ddot w}+\frac12(1+h)h'w'^3+2(1+h){\dot w}w'{\dot w}'+(1+h)^2w'^2w''\right)=0. 
\end{align}

Near the asymptotic limit $z \rightarrow 0 $, the equation \eqref{minEq} is solved order by order in $z$  by inserting the expansion
\begin{align}\label{asyExp}
&h(z)=h_0 z^\alpha+h_1 z^{2\alpha}+\cdots,\nn\\
&w(x,z)=w_0(x)+w_2(x)z^2+w_{\alpha+2}(x)z^{\alpha+2}+w_4(x)z^{4}+\cdots,
\end{align}
where we read  from \eqref{gh(z)} that $\alpha>0$ and $\alpha$ is the smaller of $2(3-\Delta)$ and $2\Delta$. 
\begin{figure}[ht!]
\centering
   \includegraphics[width = 9.5cm]{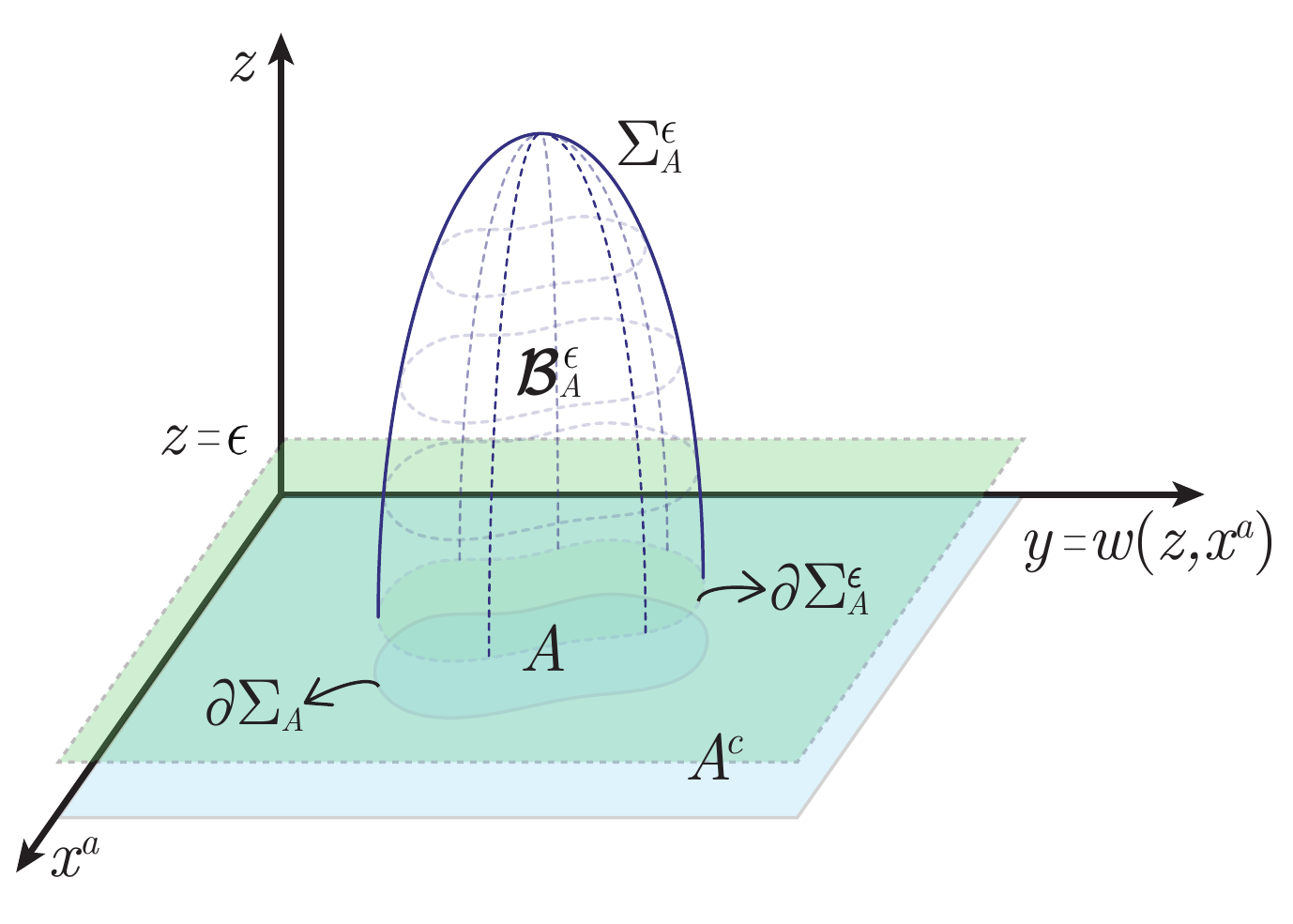}
    \caption{The RT surface and the subregion volume embedded into the asymptotically AdS$_{d+1}$ geometry.} \label{Figure1}
\end{figure}
From the leading order of \eqref{minEq}, one can determines $w_2$ in terms of $w_0$ as
\begin{align}\label{w24}
w_2=\frac{\ddot{w_0}}{2\big(1+\dot w_0^2\big)}.
\end{align}
Similarly, the higher order coefficients can also be determined, however, those are not required to obtain the gauge invariant counter term in the asymptotically AdS$_4$ geometry. Plugging the expansion \eqref{asyExp} into \eqref{Area4} and introducing a cut-off $z=\epsilon$ shown in Fig.1, we obtain the regularized HEE, 
\begin{align}\label{Area4a}
S_A^{{\rm reg}}=\frac{L^2}{4G_4}\int dx \int_\epsilon^{z_m} dz\frac{(1+h)^{1/2}}{z^2}\sqrt{1+{\dot w_0}^2}\left(1+\frac{2w_2^2+\dot w_0\dot w_2}{1+\dot w_0^2}z^2+{\cal O}(z^{\alpha+2},z^4)\right), 
\end{align} 
where $z_m$ denotes the maximum value of $z$ and is determined from the boundary condition $w'(z_m,x)\to\infty$. Using the expansion of $h(z)$ in \eqref{asyExp}, one can rewrite the regularized HEE in \eqref{Area4a} as   
\begin{align}\label{Area4b}
S_A^{{\rm reg}}=\frac{L^2}{4G_4}\int dx \int_\epsilon^{z_m} dz\sqrt{1+{\dot w_0}^2}\left(z^{-2}+\frac {h_0}2z^{\alpha-2}+{\cal O}(z^{2\alpha-2})\right).
\end{align} 
Evaluating the $z$ integral, we obtain 
\begin{align}\label{Area4c}
S_A^{{\rm reg}}= \left\{ \begin{array}{lr} \frac{L^2}{4G_4}\int dx \sqrt{1+{\dot w_0}^2}\left(\frac1\epsilon-\frac {h_0\epsilon^{\alpha-1}}{2(\alpha-1)}+{\cal O}(\epsilon^{2\alpha-1})+\cdots\right), 
& 0<\alpha<1 
\\
\frac{L^2}{4G_4}\int dx \sqrt{1+{\dot w_0}^2}\left(\frac1\epsilon-\frac {h_0}{2}\ln\Big(\frac\epsilon\ell\Big)+ \cdots \right), & \alpha=1
\\ 
\frac{L^2}{4G_4}\int dx \sqrt{1+{\dot w_0}^2}\left(\frac1\epsilon + \cdots \right), & \alpha>1
\end{array}\right.,
\end{align}
where for the $\alpha=1$ case, we have introduced an arbitrary length scale $\ell$, which  implies that the renormalized HEE will be renormalization scheme dependent. From now onwards the ellipses denote less divergent terms and finite terms.
The ranges of $\alpha$ in the above equation refers to the following ranges of the conformal dimension $\Delta$, 
\begin{align}\label{alpDel4}
0<\alpha<1 \quad  &\Longleftrightarrow \quad  0< \Delta<\frac{1}{2} \, {\rm \quad or}\quad \, \frac{5}{2}< \Delta<3,
\nn \\
\alpha=1 \quad  &\Longleftrightarrow \quad \Delta =\frac{1}{2},\,\, \frac{5}{2}, 
\nn \\
\alpha > 1   \quad  &\Longleftrightarrow \quad \frac{1}{2} < \Delta < \frac{5}{2}.
\end{align}
Based on the value of the conformal dimension $\Delta$, the value of $h_0$ in \eqref{Area4c} is determined, for instance, $h_0 = -\frac{s_0^2}{8}\sim  J_{{\cal O}^\Delta}^2 $ for $\frac{3}{2} <\Delta< 3$ and $h_0 = -\frac{v_0^2}{8}\sim  \langle {\cal O}^\Delta\rangle^2 $ for $0 <\Delta< \frac{3}{2}$, where we have used the GKP-W relations in \eqref{s0v0}.

In general, the divergences in HEE are cancelled by counter terms which are composed of the invariant quantities built by the scalar field $\phi$, the induced metric $\tilde\gamma_{ab}$, and the extrinsic curvature $\tilde K_{ab}$ on the boundary $\partial\Sigma_A^{\epsilon}$ of the regularized minimal surface 
$\Sigma_A^{\epsilon}$~\cite{Taylor:2016aoi}. See Fig.1.
In our case, the boundary $\partial \Sigma_A^{\epsilon}$ is a one dimensional curve and its embedding  into the cut-off surface $z=\epsilon$ is parametrized as $x^i=\big(x,y=w(\epsilon,x)\big)$. The induced metric and the extrinsic curvature are given by
\begin{align}\label{gammaK}
&\tilde\gamma_{xx}=\frac{L^2(1+h)}{\epsilon^2}\Big(1+{\dot w}^2\Big),\qquad \tilde K_{xx}=\frac{L(1+h)^{\frac12}\ddot w}{\epsilon(1+\dot w^2)^{5/2}},\nn\\
& \tilde K_{xy}=\frac{L(1+h)^{\frac12}\ddot w\dot w}{\epsilon(1+\dot w^2)^{5/2}},\qquad \tilde K_{yy}=\frac{L(1+h)^{\frac12}\ddot w\dot w^2}{\epsilon(1+\dot w^2)^{5/2}}.
\end{align}
The trace of the extrinsic curvature is 
\begin{align}
\tilde K=g^{ij}\tilde K_{ij}=\frac{\epsilon}{L\sqrt{1+h}}\left(\frac{\ddot w_0}{(1+\dot w_0^2)^{3/2}}+{\cal O}(\epsilon^2)\right)=\frac{\epsilon}{L\sqrt{1+h}}\left(\frac{2w_2}{(1+\dot w_0^2)^{1/2}}+{\cal O}(\epsilon^2)\right),
\end{align}
where we have used the result in \eqref{w24} in the last step. We notice that non of the divergences in \eqref{Area4c} are related to $\tilde K$. Actually, such extrinsic curvature plays a role in the renormalized  HEE if one consider the $d>3$ cases. See $d=4$ case in the next subsection. Therefore, in the case at hand the counter terms  contain only the invariants of $\phi$ and $\tilde\gamma_{ab}$.
 

The counter term required to cancel the leading divergence for $\alpha >0$ in \eqref{Area4c} is given by
\begin{align}\label{CT1-4}
S^{(1)}_{\rm ct}=-\frac{L}{4G_4}\int dx \sqrt{\tilde\gamma}=-\frac{L^2}{4G_4}\int dx\sqrt{1+{\dot w_0}^2}\left(\frac1\epsilon+\frac {h_0}2\epsilon^{\alpha-1}+{\cal O}(\epsilon^{2\alpha-1})\right).
\end{align}
Adding this counter term to \eqref{Area4c}, we obtain
\begin{align}\label{Area4d}
S_A^{{\rm reg}} +S^{(1)}_{\rm ct}= \left\{ \begin{array}{lr} \frac{L^2}{4G_4}\int dx \sqrt{1+{\dot w_0}^2}\left(-\frac {\alpha h_0\epsilon^{\alpha-1}}{2(\alpha-1)}+{\cal O}(\epsilon^{2\alpha-1})+\cdots \right), 
& 0<\alpha<1 
\\
\frac{L^2}{4G_4}\int dx \sqrt{1+{\dot w_0}^2}\left(-\frac {h_0}{2}\ln\Big(\frac\epsilon\ell\Big)+ \cdots \right), & \alpha=1
\\ 
\frac{L^2}{4G_4}\int dx \sqrt{1+{\dot w_0}^2}\Big( finite~~ terms \Big), & \alpha>1
\end{array}\right. .
\end{align}
From the solutions in \eqref{gh(z)} we note
\begin{align}
\phi^2=\phi_0^2z^\alpha+\cdots=-8h_0z^\alpha+\cdots,
\end{align}
where $\phi_0=s_0$ for the first solution and $\phi=v_0$ for the second solution in \eqref{gh(z)}.
Therefore, the counter terms that cancel the subleading divergences in \eqref{Area4d} are given by
\begin{align}\label{Sct2}
S^{(2)}_{\rm ct}= \left\{ \begin{array}{lr} -\frac{\alpha L}{64(\alpha-1)G_4}\int dx \sqrt{\tilde\gamma}\phi^2=\frac{L^2}{4G_4}\int dx \sqrt{1+{\dot w_0}^2}\left(\frac {\alpha h_0\epsilon^{\alpha-1}}{2(\alpha-1)}+\cdots \right), 
& 0<\alpha<1 
\\
-\frac{ L}{64G_4}\ln\Big(\frac\epsilon\ell\Big)\int dx \sqrt{\tilde\gamma}\phi^2=\frac{L^2}{4G_4}\int dx \sqrt{1+{\dot w_0}^2}\left(\frac {h_0}{2}\ln\Big(\frac\epsilon\ell\Big)+ \cdots \right), & \alpha=1
\end{array}\right.,
\end{align}
where we notice that the renormalized HEE for $\alpha >1$ was already obtained in \eqref{Area4d}.
For $0<\alpha<1$, there is still remaining divergence of the order $\epsilon^{2\alpha-1}$. This divergence and other less divergent terms ${\cal O}(\epsilon^{n\alpha-1})$ are cancelled by the counter terms  ${\cal A}_n\int dx \sqrt{\tilde\gamma}\phi^{2n}$. From \eqref{Area4d} and \eqref{Sct2}, we define a coordinate independent  renormalized HEE of entangling regions with the embedding \eqref{embed0} under relevant perturbations 
\begin{align}\label{Sren4}
S_A^{{\rm ren}} =S_A^{{\rm reg}}+S^{(1)}_{\rm ct} + S^{(2)}_{\rm ct}.
\end{align} 
These renormalized HEEs under relevant perturbations reproduce those for the disk entangling region in \cite{Taylor:2016aoi}.

\subsection{Renormalized HEE in asymptotically  AdS$_5$ geometry}\label{D5}
In order to test the generality of the renormalization procedure we discussed in the previous subsection, lets extend it to the case of an asymptotically AdS$_5$ geometry under relevant perturbations. Similarly with the case of the asymptotically AdS$_4$ geometry, we start from the action \eqref{Sgf}. Under assumption of the Poincar\'e invariance for the coordinate $x^\mu$,   the asymptotically AdS$_5$  metric in the Fefferman-Graham coordinate system is given by
\begin{align}
ds^2=\frac{L^2}{z^2}\Big(dz^2+\big(1+h(z)\big)\eta_{\mu\nu}dx^\mu dx^\nu\Big)\quad {\rm with} ~x^\mu=(t,x_1,x_2,y).
\end{align}
Then one can write the two independent solutions to equations of motions for the metric fluctuation and the corresponding scalar field as
\begin{align}\label{h(z)12}
&\phi_a(z)=s_0z^{4-\Delta}+s_{1}z^{3(4-\Delta)}+\cdots \Longrightarrow \quad h_a(z)=-\frac{s_0^2}{12} z^{2(4-\Delta)}+h_1z^{4(4-\Delta)}+\cdots,\nn\\
&\phi_b(z)=v_0z^{\Delta}+v_{1}z^{3\Delta} + \cdots \quad \quad~\Longrightarrow\quad h_b(z) =-\frac{v_0^2}{12}z^{2\Delta}+\tilde h_1z^{4\Delta}+\cdots.
\end{align}

The RT minimal area of the hyper-surface $\Sigma_A$ with entangling regions denoted in \eqref{embed0} for the asymptotically AdS$_5$ geometry is a three dimensional manifold parametrized by the embedding 
\begin{align}\label{embed2}
x^p=\Big(z,t=t_0,x_1,x_2, y=w(z,x_1,x_2)\Big).
\end{align}
See Fig.1. The induce metric on $\Sigma_A$ is given by
\begin{align}\label{indMet5}
&\gamma_{zz}=\frac{L^2}{z^2}\Big(1+w'^2(1+h)\Big),
\qquad \gamma_{za}=\frac{L^2}{z^2}{\partial_a w}w'(1+h),\nn\\
& \gamma_{ab}=\frac{L^2}{z^2}(1+h)\Big(\delta_{ab}+{\partial_a w}\partial_b w\Big),
\end{align}
where $\partial_a=(\partial_{x^1},\partial_{x^2}),~ w'=\partial_z w$.
Then the HEE is determined by the minimum value of the area of this hyper-surface given by 
\begin{align}\label{Area5}
{\cal A}_{\Sigma_A}=L^3\int d^2x \int dz\frac{(1+h)}{z^3}\sqrt{1+{(\partial_a w)}^2+(1+h)w'^2}. 
\end{align} 
The minimum area equation derived from this action is read as 
\begin{align}\label{minEq5}
&z(1+h)w''+z{\partial_a^2 w}-3(1+h)w'+2zh'w'\nn\\
-&\frac z{{1+{\partial_c w}^2+(1+h)w'^2}}\left[{\partial_a w\partial_b w}{\partial_a\partial_b w}+(1+h)^2\Big(w'^2w''+\frac{\frac12h'w'^3+2w'{\partial_a w}{\partial_a w'}}{1+h}\Big)\right]=0. 
\end{align} 
We introduce the asymptotic expansion of $w(z,x_1,x_2)$ and solve this equation order by order in $z$. In this case, which is the true in any case of even boundary dimension $d$,  the iteration breaks down at $z^{d=4}$ order, and one needs to introduce logarithmic term at this order. We can also read the asymptotic expansion of  $h(z)$ from \eqref{h(z)12}, 
\begin{align}\label{asym5}
h(z) &= h_0z^{\alpha}+h_1z^{2\alpha}+\cdots,\nn\\
w(z,x^a)&= \left\{ \begin{array}{lr} w_0(x^a)+w_2(x^a)z^2+w_{\alpha+2}(x^a)z^{\alpha+2}+\cdots, 
& 0<\alpha<2 
\\
w_0(x^a)+w_2(x^a)z^2+w_{4}(x^a)z^{4}+\tilde w_{4}(x^a)z^{4}\ln(z)+\cdots, & \alpha\ge2
\end{array}\right.,
\end{align}
where $\alpha$ is the smaller of $2(4-\Delta)$ and $2\Delta$, and then $\alpha$ and $\Delta$ have the relations
\begin{align}\label{alpDel5}
0<\alpha<2 \quad  &\Longleftrightarrow \quad  0< \Delta<1 \, {\rm \quad or}\quad \, 3< \Delta<4,
\nn \\
\alpha=2 \quad  &\Longleftrightarrow \quad \Delta = 1, 3, 
\nn \\
\alpha > 2   \quad  &\Longleftrightarrow \quad 1 < \Delta < 3.
\end{align}
The $w_n$'s in \eqref{asym5} are determined in terms of $w_0$ by solving the minimal area equation \eqref{minEq5}.  For our purpose in this and the next subsections, we only need $w_2$ and $w_{\alpha+2}\,(0<\alpha<2)$. They are expressed as
\begin{align}\label{w2}
w_2=\frac14\left(\partial_a^2w_0-\frac{\partial_a w_0\partial_b w_0\partial_a\partial_b w_0}{1+(\partial_cw_0)^2}\right),\qquad w_{\alpha+2}= -\frac{4 h_0 (\alpha -1) w_2}{\alpha ^2-4}.
\end{align}

Plugging the expansion \eqref{asym5} into \eqref{Area5} and introducing the cut-off surface $z=\epsilon$, the HEE is given by
\begin{align}\label{S5a}
S_A^{{\rm reg}}=\frac{L^3}{4G_5}\int d^2x \int_\epsilon^{z_m} dz\frac{(1+h)\sqrt{1+(\partial_c w_0)^2}}{z^3}\left[1+\frac{2w_2^2+\partial_a w_0\partial_a w_2}{1+(\partial_c w_0)^2}z^2+{\cal O}(z^{\alpha+2})\right]. 
\end{align} 
Integrating by parts the second term in the square bracket in \eqref{S5a} and then using the result in \eqref{w2}, we obtain
\begin{align}\label{S5a-1}
S_A^{{\rm reg}}=\frac{L^3}{4G_5}\int d^2x \int_\epsilon^{z_m} dz\sqrt{1+(\partial_c w_0)^2}\left[\frac1{z^{3}}-\frac{2w_2^2}{1+(\partial_c w_0)^2}\frac1z+h_0z^{\alpha-3}+{\cal O}(z^{\alpha-1})\right], \end{align} 
where $h_0 = -\frac{s_0^2}{12}\sim J_{{\cal O}^\Delta}^2$ for $2<\Delta<4$ and $h_0 =  -\frac{v_0^2}{12}\sim \langle {\cal O}^\Delta\rangle^2$ for $0<\Delta<2$, and $w_2$ is expressed in terms of $w_0$ in \eqref{w2}. 
We carry out the $z$ integration and  obtain
\begin{align}\label{S5b}
S_A^{{\rm reg}} = \left\{ \begin{array}{lr} \frac{L^3}{4G_5}\int d^2x\sqrt{1+(\partial_c w_0)^2}\left[\frac1{2\epsilon^{2}}+\frac{2w_2^2}{1+(\partial_c w_0)^2}\ln\Big(\frac\epsilon\ell\Big)-\frac{h_0\epsilon^{\alpha-2}}{\alpha-2}+\cdots\right], 
& 0<\alpha<2 
\\
\frac{L^3}{4G_5}\int d^2x\sqrt{1+(\partial_c w_0)^2}\left[\frac1{2\epsilon^{2}}+\frac{2w_2^2}{1+(\partial_c w_0)^2}\ln\Big(\frac\epsilon\ell\Big)-{h_0}\ln\Big(\frac\epsilon\ell\Big)+\cdots\right], & \alpha=2
\\ 
\frac{L^3}{4G_5}\int d^2x\sqrt{1+(\partial_c w_0)^2}\left[\frac1{2\epsilon^{2}}+\frac{2w_2^2}{1+(\partial_c w_0)^2}\ln\Big(\frac\epsilon\ell\Big)+\cdots\right], & \alpha>2
\end{array}\right.,
\end{align}
where we also introduce some length scale $\ell$ like the case of the asymptotically AdS$_4$ in the previous subsection.
Next we fix the counter terms which are composed of the invariant quantities of the scalar field $\phi$,  the induced metric, and extrinsic curvature on the boundary space $\partial\Sigma_A^{\epsilon}$. In this case the boundary space is a 2-dimensional surface embedded in the cut of surface $z=\epsilon$ as
$x^i=(x_1,x_2,y=w(\epsilon,x_1,x_2))$. The induced metric and the extrinsic curvature are obtained as
\begin{align}
&\tilde\gamma_{ab}=\frac{L^2}{\epsilon^2}\big(1+h\big)\big(\delta_{ab}+\partial_aw\partial_bw\big),\nn\\
&\tilde K_{ab}=\frac{L(1+h)^{1/2}}{\epsilon\sqrt{1+(\partial_cw)^2}}\left(\partial_a\partial_b w-\frac{\partial_aw\partial_b\partial_dw\partial_d w+\partial_bw\partial_a\partial_dw\partial_d w}{1+(\partial_cw)^2}+\frac{\partial_aw\partial_bw\partial_dw\partial_ew\partial_d\partial_ew}{(1+(\partial_cw)^2)^2}\right),\nn\\
&\tilde K_{ay}=\frac{L(1+h)^{1/2}}{\epsilon\sqrt{1+(\partial_cw)^2}}\left(\frac{\partial_a\partial_b\partial_bw}{1+(\partial_cw)^2}-\frac{\partial_aw\partial_bw\partial_dw\partial_b\partial_dw}{(1+(\partial_cw)^2)^2}\right),
\nn\\
&\tilde K_{yy}=\frac{L(1+h)^{1/2}}{\epsilon\sqrt{1+(\partial_cw)^2}}\left(\frac{\partial_aw\partial_bw\partial_a\partial_bw}{(1+(\partial_cw)^2)^2}\right).
\end{align}
Using the asymptotic expansion in \eqref{asym5}, the trace of the extrinsic curvature is
given by\begin{align}\label{TrKt}
\tilde K=\frac{\epsilon(1+h)^{-1/2}}{L\sqrt{1+(\partial_cw_0)^2}}\left(\partial^2_aw_0-\frac{\partial_aw_0\partial_bw_0\partial_a\partial_bw_0}{1+(\partial_cw_0)^2}\right)+{\cal O}(\epsilon^3)=\frac{4\epsilon(1+h)^{-1/2}}{L\sqrt{1+(\partial_cw_0)^2}}w_2+{\cal O}(\epsilon^3),
\end{align}
where in the last step we have used the expression of $w_2$ in \eqref{w2}.

The first counter term that cancels the leading order divergence in \eqref{S5b} is 
\begin{align}\label{Sct1}
S^{(1)}_{\rm ct}=-\frac{L}{8G_5}\int d^2x\sqrt{\tilde \gamma}=-\frac{L^3}{4G_5}\int d^2x\sqrt{1+(\partial_c w_0)^2}\left[\frac1{2\epsilon^{2}}+\frac{h_0}2\epsilon^{\alpha-2}+\cdots\right].
\end{align}
Adding the counter term \eqref{Sct1} to \eqref{S5b}, we obtain
\begin{align}\label{S5c}
S_A^{{\rm reg}}+S^{(1)}_{\rm ct} = \left\{ \begin{array}{lr} \frac{L^3}{4G_5}\int d^2x\sqrt{1+(\partial_c w_0)^2}\left[\frac{2w_2^2}{1+(\partial_c w_0)^2}\ln\Big(\frac\epsilon\ell\Big)-\frac{\alpha h_0\epsilon^{\alpha-2}}{2(\alpha-2)}+\cdots\right], 
& 0<\alpha<2 
\\
\frac{L^3}{4G_5}\int d^2x\sqrt{1+(\partial_c w_0)^2}\left[\frac{2w_2^2}{1+(\partial_c w_0)^2}\ln\Big(\frac\epsilon\ell\Big)-{h_0}\ln\Big(\frac\epsilon\ell\Big)+\cdots\right], & \alpha=2
\\ 
\frac{L^3}{4G_5}\int d^2x\sqrt{1+(\partial_c w_0)^2}\left[\frac{2w_2^2}{1+(\partial_c w_0)^2}\ln\Big(\frac\epsilon\ell\Big)+\cdots\right], & \alpha>2
\end{array}\right..
\end{align}
The universal logarithmic divergences which are present in the all the three ranges of $\alpha$ are cancelled by the counter term composed of trace of the extrinsic curvature in \eqref{TrKt}. As we mentioned in the previous subsection this term is absent for $d<4$. The required counter term is\footnote{In \cite{Taylor:2016aoi}, it was argued that odd powers of $\tilde K$ can not enter the counter terms because the renormalized HEE for the subspace $A$ and its complement $A^c$, which has opposite sign of the extrinsic curvature,  must be the same.} 
\begin{align}\label{Sct2-1}
S^{(2)}_{\rm ct}=-\frac{L^3}{32G_5}\ln\Big(\frac\epsilon\ell\Big)\int d^2x\sqrt{\tilde \gamma}\tilde K^2=-\frac{L^3}{4G_5}\ln\Big(\frac\epsilon\ell\Big)\int d^2x\left[\frac{2w_2^2}{\sqrt{1+(\partial_c w_0)^2}}+{\cal O}({\epsilon^2})+\cdots\right].
\end{align}
Then we obtain
\begin{align}\label{S5d}
S_A^{{\rm reg}}+S^{(1)}_{\rm ct}+S^{(2)}_{\rm ct}= \left\{ \begin{array}{lr} \frac{L^3}{4G_5}\int d^2x\sqrt{1+(\partial_c w_0)^2}\left[-\frac{\alpha h_0\epsilon^{\alpha-2}}{2(\alpha-2)}+\cdots\right], 
& 0<\alpha<2 
\\
\frac{L^3}{4G_5}\int d^2x\sqrt{1+(\partial_c w_0)^2}\left[-{h_0}\ln\Big(\frac\epsilon\ell\Big)+\cdots\right], & \alpha=2
\\ 
\frac{L^3}{4G_5}\int d^2x\sqrt{1+(\partial_c w_0)^2}\left(finite\,\, terms\right), & \alpha>2
\end{array}\right..
\end{align}
The third counter term to cancel the remaining two divergences is composed of  the scalar field $(\phi^2=-12h_0z^\alpha+\cdots)$,
\begin{align}\label{Sct3}
S^{(3)}_{\rm ct}= \left\{ \begin{array}{lr} -\frac{\alpha L}{96(\alpha-2)G_5}\int d^2x\sqrt{\tilde \gamma}\,\phi^2, 
& 0<\alpha<2 
\\
-\frac{L}{96G_5}\ln\Big(\frac\epsilon\ell\Big)\int d^2x\sqrt{\tilde \gamma}\,\phi^2, & \alpha=2
\end{array}\right..
\end{align}
This counter term cancels all the divergences except for $\alpha<1$ where there are less divergent terms ${\cal O}(\epsilon^{2\alpha-2})$, which can be removed by counter terms containing higher powers of $\phi$ and $\tilde K$.
Then from \eqref{S5d} and \eqref{Sct2}, one can define the renormalized HEEs under relevant perturbations for entangling regions in the asymptotically AdS$_5$ geometry.

\section{Renormalized  HSC  under Relevant Perturbations}
\label{RHSC} 

The CV conjecture for the subregion 
complexity~\cite{Alishahiha:2015rta} states that the HSC is equal to the volume of the codimension-one hypersurface  ${\cal B}_A$ enclosed by the entangling subregion $A$ and the corresponding RT surface $\Sigma_A$~\cite{Ryu:2006bv}, i.e., 
\begin{align}\label{CA1}
{\cal C}_A = \frac{V({\cal B}_A)}{8\pi L G_{d+1}},
\end{align}
where $L$ is the radius of the AdS$_{d+1}$ geometry. 
 The CA conjecture for the subregion complexity was also proposed in \cite{Carmi:2016wjl}.
In this section, we construct the renormalized HSC with entangling regions denoted by \eqref{embed0}  for the asymptotically AdS$_{4,5}$ geometries under relevant perturbations. To do that, one has to consider divergent terms generated from the boundary of the entangling region $\partial A$, which is  different from the renormalization of the holographic complexity for the whole space~\cite{Kim:2017lrw}.

\subsection{Renormalized HSC in asymptotically AdS$_4$ geometry}\label{RHSC4} 

According to  proposal of the HSC~\cite{Alishahiha:2015rta}, we use the RT surface $\Sigma_A$ for entangling regions denoted by \eqref{embed0}, which were obtained in the previous section.  
To regularize divergences of the HSC ${\cal C}_A$ in \eqref{CA1}, we also introduce  the $z=\epsilon$ cut-off, and then the regularized HSC is written as  
 \begin{align}\label{HSC2}
{\cal C}_{A}^{{\rm reg}} =\frac1{8\pi LG_4}\int dx \int_\epsilon^{z_m} dz \int_0^{w(z,x)} dy\frac{L^3}{z^3}(1+h)=\frac{L^2}{8\pi G_4}\int dx \int_\epsilon^{z_m} dz  \frac{w(z,x)}{z^3}(1+h),
\end{align}
where we used the asymptotically AdS$_4$  geometry in \eqref{FG1} and $w(z,x)$ is defined in the embedding \eqref{wofzx}. 
Using the asymptotic expansion \eqref{asyExp}, we obtain
\begin{align}\label{HSC3}
{\cal C}_{A}^{{\rm reg}}=\frac{L^2}{8\pi G_4}\int dx \int_\epsilon^{z_m} dz  \Big(w_0z^{-3}+w_2z^{-1}+h_0 w_0z^{\alpha-3}+{\cal O}(\epsilon^{2\alpha-3})+\cdots+{\cal O}(z^{\alpha-1})\Big).
\end{align}
Evaluating the $z$ integral, we  single out the divergent terms as follows:
\begin{align}\label{HSC4}
{\cal C}_{A}^{{\rm reg}}= \left\{ \begin{array}{lr} \frac{L^2}{8\pi G_4}\int dx \bigg(\frac{w_0}{2\epsilon^2}-w_2\ln\Big(\frac\epsilon \ell\Big)+\frac{h_0 w_0\epsilon^{\alpha-2}}{(2-\alpha)}+\frac{h_1w_0\epsilon^{2\alpha-2}}{2(1-\alpha)}+\cdots\bigg),
& 0<\alpha<2 
\\
\frac{L^2}{8\pi G_4}\int dx \left(\frac{w_0}{2\epsilon^2}-\big(w_2+h_0 w_0\big)\ln\Big(\frac\epsilon\ell\Big)+\cdots\right), & \alpha=2
\\ 
\frac{L^2}{8\pi G_4}\int dx \left(\frac{w_0}{2\epsilon^2}-w_2\ln\Big(\frac\epsilon\ell\Big)+\cdots\right), & \alpha>2
\end{array}\right.,
\end{align}
where the relations between $\alpha$ and the conformal dimension $\Delta$ were given in \eqref{alpDel4}. For $\alpha=1$ the last divergence in the first line will be logarithmic. We note that in the case of $\alpha=2$, there is only one divergence that depends on the scalar deformation whereas those divergences are absent when $\alpha>2$.

For a $(d+1)$-dimensional bulk space-time, the counter terms are given by
\begin{align}\label{CTG}
{\cal C}_{\rm ct}=\int_{\partial {\cal M}_\epsilon}d^{d-1}x\sqrt{\det h_{ij}}\sum_{n}{\cal C}_{n}(g_{\mu\nu},R_{\mu\nu},h_{ij},K_{ij}),
\end{align}
where $\partial {\cal M}_\epsilon$ is a codimension-two static hyper-surface at the cut-off boundary $(z=\epsilon)$,   $h_{ij}$ is the induced metric  on $\partial {\cal M}_\epsilon$,  and $K_{ij}$ is the extrinsic curvature of the $\partial {\cal M}_\epsilon$ embedded in the bulk constant time slice ${\cal M}_\epsilon$. In addition, 
$g_{\mu\nu}$ is the induced metric on the cut-off $z=\epsilon$ boundary, $R_{\mu\nu}$  is the Ricci tensor derived from $g_{\mu\nu}$, and ${\cal C}_{n}$ are invariants built from $R_{\mu\nu},g_{\mu\nu},h_{ij}$, and $K_{ij}$ with appropriate mass dimensions. 
In the case we are considering, both $g_{\mu\nu}$ and  $h_{ij}$ at $z=\epsilon$ are flat, i.e.,
\begin{align}
g_{\mu\nu}=\frac{L^2}{\epsilon^2}\big[1+h(\epsilon)\big]\eta_{\mu\nu},\quad\qquad h_{ij}=\frac{L^2}{\epsilon^2}\big[1+h(\epsilon)\big]\delta_{ij},
\end{align}
where $\mu,\nu=(t,x,y)$ and $i,j=(x,y)$.  Therefore, the Ricci tensor $R_{\mu\nu}$  is vanishing and the extrinsic curvature is given by 
\begin{align}\label{Kij}
K_{ij}=&L\left(\frac{h'(\epsilon)}{2\epsilon}-\frac{1+h(\epsilon)}{\epsilon^2}\right)\delta_{ij},\qquad K_{iz}=0,\qquad K_{zz}=0.
\end{align} 
 
According to the general formula in \eqref{CTG}, the counter term which cancels the leading order divergence in \eqref{HSC4} is given by
\begin{align}
{\cal C}_{\rm ct}^{(1)}=-\frac{1}{16\pi G_4}\int dx \int_{0}^{w(\epsilon, x)}dy\sqrt{\det h_{ij}}=-\frac{L^2}{16\pi G_4}\int dx\left(\frac{w_0}{\epsilon^2}+\frac{h_0 w_0}{\epsilon^{2-\alpha}}+\frac{h_1w_0}{\epsilon^{2-2\alpha}}+\cdots\right). 
\end{align}
Adding this counter term to the HSC in \eqref{HSC4}, we obtain
\begin{align}\label{HSC5}
{\cal C}_{A}^{{\rm reg}}+{\cal C}_{\rm ct}^{(1)}= \left\{ \begin{array}{lr} \frac{L^2}{8\pi G_4}\int dx \bigg(-w_2\ln\Big(\frac\epsilon \ell\Big)+\frac{\alpha h_0 w_0\epsilon^{\alpha-2}}{2(2-\alpha)}+\frac{\alpha h_1w_0\epsilon^{2\alpha-2}}{2(1-\alpha)}+\cdots\bigg),
& 0<\alpha<2 
\\
\frac{L^2}{8\pi G_4}\int dx \left(-w_2\ln\Big(\frac\epsilon\ell\Big)-h_0 w_0\ln\Big(\frac\epsilon\ell\Big)+\cdots\right), & \alpha=2
\\ 
\frac{L^2}{8\pi G_4}\int dx \left(-w_2\ln\Big(\frac\epsilon\ell\Big)+\cdots\right), & \alpha>2
\end{array}\right..
\end{align}
The universal $w_2\ln(\epsilon/\ell)$ divergence is independent of the relevant perturbations by the scalar field and it is  present in the case of pure AdS$_4$ background as well. We will come back to renormalization of this term later. First, let us discuss the other divergences. Since the Ricci tensor $R_{\mu\nu}$ is vanishing, the counter terms  to cancel these other divergences are built from the invariants of $ K_{ij}$ and $h_{\ij}$.
The counter term that has the right structure to cancel the second terms in the first and second lines of \eqref{HSC5}  is  
\begin{align}
{\cal C}_{\rm ct}^{(2)}= \left\{ \begin{array}{lr} \frac{1}{8\pi G_4(\alpha-2)}\int dx \int_{0}^{w(\epsilon,   x)}dy\sqrt{\det h_{ij}}\big(1+\frac{LK}{2}\big), 
& 0<\alpha<2 
\\
\frac{1}{8\pi G_4}\ln\Big(\frac\epsilon\ell\Big)\int dx \int_{0}^{w(\epsilon,x)}dy\sqrt{\det h_{ij}}\big(1+\frac{LK}{2}\big), & \alpha=2
\end{array}\right..
\end{align}
Then we obtain 
\begin{align}\label{HSC6}
{\cal C}_{A}^{{\rm reg}}+{\cal C}_{\rm ct}^{(1)}+{\cal C}_{\rm ct}^{(2)}= \left\{ \begin{array}{lr} \frac{L^2}{8\pi G_4}\int dx \bigg(-w_2\ln\Big(\frac\epsilon \ell\Big)+\frac{\alpha^2h_1w_0\epsilon^{2\alpha-2}}{2(\alpha-2)(\alpha-1)}+\cdots\bigg),
& 0<\alpha<2 
\\
\frac{L^2}{8\pi G_4}\int dx \left(-w_2\ln\Big(\frac\epsilon\ell\Big)+\cdots\right), & \alpha=2
\\ 
\frac{L^2}{8\pi G_4}\int dx \left(-w_2\ln\Big(\frac\epsilon\ell\Big)+\cdots\right), & \alpha>2
\end{array}\right..
\end{align}
From \eqref{HSC6}, we see that  there is a divergence which is of order $\epsilon^{2\alpha-2}$ for $\alpha<1$.  In order to cancel this divergence we need to add counter terms that contain  $K^2$. Note that $K^{ij}K_{ij}$ also has the right order of divergence, however, since in our case $K^{ij}K_{ij}=\frac12 K^2$, it is not an independent contribution. Therefore, the required counter term at this order is 
\begin{align}\label{Cct3}
{\cal C}_{\rm ct}^{(3)}=&-\frac{2h_1}{8\pi G_4(\alpha-2)(\alpha-1)h_0^2}\int dx \int_{0}^{w(\epsilon, x)}dy\sqrt{\det h_{ij}}\big(1+L K+\frac{L^2}{4} K^2\big),\quad &0<\alpha<1.
\end{align}
Adding the counter term \eqref{Cct3} to \eqref{HSC6}, we obtain \begin{align}\label{HSC7}
{\cal C}_{A}^{{\rm reg}}+{\cal C}_{\rm ct}^{(1)}+{\cal C}_{\rm ct}^{(2)}+{\cal C}_{\rm ct}^{(3)}= \left\{ \begin{array}{lr} \frac{L^2}{8\pi G_4}\int dx \bigg(-w_2\ln\Big(\frac\epsilon \ell\Big)+{\cal O}(\epsilon^{3\alpha-2})+\cdots\bigg),
& 0<\alpha<2 
\\
\frac{L^2}{8\pi G_4}\int dx \left(-w_2\ln\Big(\frac\epsilon\ell\Big)+\cdots\right), & \alpha=2
\\ 
\frac{L^2}{8\pi G_4}\int dx \left(-w_2\ln\Big(\frac\epsilon\ell\Big)+\cdots\right), & \alpha>2
\end{array}\right..
\end{align}
Similarly, less  divergent terms which are ${\cal O}(\epsilon^{n\alpha-2})$ are removed by adding counter terms that are higher order in $K$.  However, the $w_2\ln(\epsilon/\ell)$ term, which is independent of the scalar deformation and it exists for all the three ranges of $\alpha$, can not be cancelled by any counter terms built from the invariants  of $R_{\mu\nu},g_{\mu\nu},h_{ij}$ $K_{ij}$. Therefore, the general form of the counter terms \eqref{CTG} proposed in the literature does not account for this particular divergence. It turns out, these divergences  are cancelled by counter terms which are built from the invariants of the induced metric $\tilde\gamma_{ab}$ and  the extrinsic curvature $\tilde K_{ab}$ on the boundary $\partial\Sigma_A^{\epsilon}$ of the regularized minimal surface 
$\Sigma_A^{\epsilon}$~\cite{Carmi:2016wjl}. 
In our case  these induced metric and the extrinsic curvature are as in \eqref{gammaK}.
The required counter term which cancels the logarithmic divergence in \eqref{HSC7} is 
 \begin{align}\label{ctct4}
\tilde {\cal C}_{\rm ct}&=\frac{L^2}{16\pi G_4}\ln\Big(\frac\epsilon\ell\Big)\int dx\sqrt{\det\tilde\gamma_{xx}}\tilde K=\frac{L^2}{8\pi G_4}\ln\Big(\frac\epsilon\ell\Big)\int dx\Big(w_2+{\cal O}(\epsilon^2)\Big).
\end{align}
 Adding this counter term to \eqref{HSC7}, we finally obtain a finite result for $\alpha\ge\frac23$ whereas for $\alpha<\frac23$, as it was stated above, we need more counter terms containing higher powers of $K$. However, we note that the number of the necessary counter terms are finite once one fixes the value of $\alpha$, therefore, in the case of asymptotically AdS$_4$ geometry, HSC is renormalizable for any value of $\alpha$.
\subsection{Renormalized HSC in asymptotically AdS$_5$ geometry}\label{RHSC5} 
The renormalization of HSC in asymptotically AdS$_5$ geometry follows the same steps as the $d=3$ case, however, in the current case there is an extra ${\cal O}(\epsilon^{\alpha-1})$ divergence which can not be cancelled by any of the counter terms listed in the previous subsection. The regularized HSC  is given by
 \begin{align}\label{HSC5-1}
{\cal C}_{A}^{{\rm reg}}=\frac{L^3}{8\pi G_5}\int d^2x \int_\epsilon^{z_m} dz (1+h)^{\frac32} \frac{w(z,x_1,x_2)}{z^4}.
\end{align}
Evaluating the $z$ integration in \eqref{HSC5-1}, we obtain
\begin{align}\label{HSC5-2}
{\cal C}_{A}^{{\rm reg}}= \left\{ \begin{array}{lr} \frac{L^3}{8\pi G_5}\int d^2x \Big(\frac{w_0}{3\epsilon^{3}}+\frac{w_2}{\epsilon}+\frac3{2(3-\alpha)} \frac{w_0h_0}{\epsilon^{3-\alpha}} +\frac{3 h_0w_2+2w_{\alpha+2}}{2(1-\alpha)\epsilon^{1-\alpha}}+{\cal O}(\epsilon^{2\alpha-3})+\cdots\Big),
&  0<\alpha<2 
\\ 
\frac{L^3}{8\pi G_5}\int d^2x \left(\frac{w_0}{3\epsilon^{3}}+\frac{w_2}{\epsilon}+\frac3{2(3-\alpha)}\frac{w_0h_0}{\epsilon^{3-\alpha}}+\cdots\right), & \alpha\ge2
\end{array}\right.,
\end{align}
where the relations between $\alpha$ and the conformal dimension $\Delta$ were given in \eqref{alpDel5}.
Note that for $\alpha=1$ and $\alpha=3$ cases, we get logarithmic divergences and those cases should be treated separately as in the previous subsection. 

As in the previous subsection, the Ricci tensor derived from the flat boundary metric $g_{\mu\nu}=\frac{L^2}{\epsilon^2}\big[1+h(\epsilon)\big]\eta_{\mu\nu}$ is vanishing.  Therefore, the counter terms are obtained from the induced metric $h_{ij}$ and the extrinsic curvature $K_{ij}$ on the cut-off surface $z=\epsilon$, which are given by
\begin{align}
h_{ij}=\frac{L^2}{\epsilon^2}\big[1+h(\epsilon)\big]\delta_{ij},\qquad K_{ij}=\frac{L}{2 \epsilon^2}{ \big[\epsilon h'(\epsilon)-2 h(\epsilon)-2\big]}\delta_{ij},
\end{align}
where $i,j=(x_1,x_2,y)$ are the coordinates on the cut-off surface at fixed time $t=t_0$. The appropriate counter term which removes the leading divergence in \eqref{HSC5-2} is 
\begin{align}
{\cal C}_{\rm ct}^{(1)}&=-\frac{1}{24\pi G_5}\int d^2x\sqrt{\det h_{ij}}~{w(\epsilon, x_1,x_2)}.
\end{align} 
Adding this counter term to \eqref{HSC5-2}, we obtain 
\begin{align}\label{HSC5-3}
{\cal C}_{A}^{{\rm reg}}+{\cal C}_{\rm ct}^{(1)} &= \left\{ \begin{array}{lr} \frac{L^3}{8\pi G_5}\int d^2x \Big(\frac{2w_2}{3\epsilon} + \frac\alpha{2(3-\alpha)}\frac{w_0h_0}{\epsilon^{3-\alpha}} + \frac{(\alpha +2)   (3 h_0 w_2+2 w_{\alpha +2})}{6 (1-\alpha) \epsilon^{1-\alpha}}+\cdots\Big),
&  0<\alpha<2 
\\ 
\frac{L^3}{8\pi G_5}\int d^2x \left(\frac{2w_2}{3\epsilon} + \frac\alpha{2(3-\alpha)}\frac{w_0h_0}{\epsilon^{3-\alpha}} +\cdots\right), & \alpha\ge2
\end{array}\right..
\end{align}
 The counter term that cancels ${\cal O}(\epsilon^{\alpha-3})$ divergence is linear in $K$ and is given by 
 \begin{align}
{\cal C}_{\rm ct}^{(2)}=&\frac{1}{8\pi G_4 (\alpha-3)}\int dx \int_{0}^{w(\epsilon, x)}dy\sqrt{\det h_{ij}}\big(1+\frac{ LK}{3}\big).
\end{align} 
Adding this counter term to \eqref{HSC5-3}, we obtain 
\begin{align}\label{HSC5-4}
{\cal C}_{A}^{{\rm reg}}+{\cal C}_{\rm ct}^{(1)}+{\cal C}_{\rm ct}^{(2)} &= \left\{ \begin{array}{lr}  \frac{L^3}{8\pi G_5}\int d^2x \left(\frac{2w_2}{3\epsilon}+\frac{9h_0 w_2+ (6+\alpha -\alpha^2)  w_{\alpha+2}}{3 (\alpha -3) (\alpha -1)\epsilon^{1-\alpha}}+\cdots\right),
&  0<\alpha<1, 
\\ 
\frac{L^3}{8\pi G_5}\int d^2x \left(\frac{2w_2}{3\epsilon}+{\rm \cdots}\right), & \alpha>1
\end{array}\right.,
\end{align}
where we notice that the ranges of $\alpha$ are changed from those of \eqref{HSC5-3} and the corresponding ranges of the conformal dimension $\Delta$ are 
\begin{align}
0<\alpha<1\qquad &\Longleftrightarrow \qquad 0< \Delta < \frac{1}{2}\quad {\rm and} \quad \frac{7}{2}< \Delta < 4, 
\nonumber\\
\alpha>1 \qquad &\Longleftrightarrow \qquad \frac{1}{2} < \Delta < \frac{7}{2}.
\end{align}

 The peculiar divergence, which is independent from the scalar field deformation, is now ${\cal O}(\epsilon^{-1})$. In order to cancel this divergence we need to introduce the extrinsic curvature on the boundary $\partial\Sigma_A^{\epsilon}$ of the RT surface, as we did in the previous subsection. This corresponding extrinsic curvature was given in \eqref{TrKt}. Therefore, the counter term that cancels the ${\cal O}(\epsilon^{-1})$ divergence in \eqref{HSC5-4} is given by
\begin{align}
\tilde {\cal C}_{\rm ct}=-\frac{L^2}{48\pi G_5}\int d^2x\sqrt{\det\tilde\gamma_{ab}}\tilde K=-\frac{L^3}{8\pi G_5}\int d^2x\left(\frac{2w_2}{3\epsilon}+\frac{h_0w_2}{3\epsilon^{1-\alpha}}+{\cal O}(\epsilon)\right).
\end{align}
Adding this counter term to \eqref{HSC5-4},  we finally obtain
 \begin{align}\label{HSC5-6}
{\cal C}_{A}^{{\rm reg}}+{\cal C}_{\rm ct}^{(1,2)}+\tilde {\cal C}_{\rm ct} &= \left\{ \begin{array}{lr}  \frac{L^3}{8\pi G_5}\int d^2x \left(\frac{\left(6+4\alpha- \alpha ^2\right) h_0 w_2+ (6+\alpha -\alpha^2) w_{\alpha+2}}{3 (\alpha -3) (\alpha -1)\epsilon^{1-\alpha}}+\cdots\right),
&  0<\alpha<1, 
\\ 
\frac{L^3}{8\pi G_5}\int d^2x \Big({\cal O} (\epsilon^{2\alpha-3})+\cdots\Big), & \alpha>1
\end{array}\right..
\end{align}
Similar to the case of the asymptotically AdS$_4$ geometry in the previous subsection, the less divergent term ${\cal O} (\epsilon^{2\alpha-3})$ in \eqref{HSC5-6} can be cancelled by adding higher order of extrinsic curvature terms, though we omit the detailed form of the counter term. However, we find that one cannot cancel the divergent term for $0<\alpha<1$ in \eqref{HSC5-6}. That is, adding a new counter term produces another divergent term, which can not be canceled out.   This implies that there is no renormalized HSC in the range $0<\alpha<1$, i.e., $0< \Delta < \frac{1}{2}\,\, {\rm and} \,\, \frac{7}{2}< \Delta < 4$. Here the latter case does not violate the unitary bound $(\Delta \ge1)$ for primary operators in 4-dimensional dual field theory.  It will be interesting if one figures out the physical reason of this phenomenon by investigating other HSC conjectures or spacetime dimensions.

\section{An Example: The LLM geometry }
In this section, we test the general procedure we discussed in the previous sections, by using an asymptotically AdS$_4$ geometry, which is obtained from the KK reduction of the 11-dimensional LLM solutions~\cite{Lin:2004nb}.  For definiteness, we choose the boundary subspace $A$ to be a disc of radius $R$. 

\subsection{The LLM Geometry }
We have presented a detailed account of the LLM geometries with (or without) discrete torsion and applied the KK holography procedure~\cite{Skenderis:2006uy,Skenderis:2006di} to obtain the vacuum expectation values (vevs) of chiral primary operators (CPOs) with conformal dimensions $\Delta = 1,2$ in the U$_k(N)\times$U$_{-k}(N)$ mass-deformed ABJM (mABJM) theory~\cite{Hosomichi:2008jb,Gomis:2008vc}.\footnote{The gauge/gravity duality between the ${\cal N} = 6$ supersymmetry preserving mABJM theory and the LLM geometry was investigated in the large $N$ limit~\cite{Jang:2016aug,Jang:2017gwd, Jang:2019pve}.}  Here the mABJM theory is obtained from the supersymmetry preserving mass deformation of the ${\cal N} = 6$ ABJM theory~\cite{Aharony:2008ug}.  In this subsection, we briefly review
some necessary aspects of the LLM geometries  and the KK reduction to asymptotically AdS$_4$ geometry. 

 The LLM geometries with SO(2,1)$\times {\rm SO}(4)/{\mathbb Z}_k \times {\rm SO}(4)/{\mathbb Z}_k$ isometry are BPS solutions of the 11-dimensional supergravity~\cite{Lin:2004nb,Cheon:2011gv }. The metric and the corresponding 4-form field strength are given by
\begin{align}\label{LLMmetric}
 ds^2 &= -G_{tt} ( -dt^2 + dw_1^2 + dw_2^2) + G_{xx} (d\tilde x^2 + d\tilde y^2) 
+ G_{\theta\theta} ds^2_{S^3/\mathbb{Z}_k}
     G_{\tilde\theta\tilde\theta} ds^2_{\tilde S^3/\mathbb{Z}_k},
\\
\label{LLMF4}
F_4 &= -d \left(e^{2\Phi}h^{-2}V \right)
\wedge dt\wedge dw_1\wedge dw_2 +\mu_0^{-1} \left[Vd(\tilde y^2e^{2G}) + h^2e^{3G}\star_2 d(\tilde y^2 e^{-2G})\right]
\wedge d\Omega_3
\nn \\
&~~~\,+\mu_0^{-1}
\left[ Vd(\tilde y^2e^{-2G}) -h^2e^{-3G}\star_2 d(\tilde y^2 e^{2G})\right]
\wedge d\tilde\Omega_3,
\end{align}
where $\mu_0$ is a mass parameter, $ds^2_{S^3/\mathbb{Z}_k}$ and $ds^2_{\tilde S^3/\mathbb{Z}_k}$ are metrics of two $S^3$'s with $\mathbb{Z}_k$ orbifold,  while $d\Omega_3$ and $d\tilde\Omega_3$ are the corresponding volume forms.
The  metric $G_{pq}$ and the 4-form field strength $F_{pqrs}$ are completely determined by the two functions $Z(\tilde x,\tilde y)$ and $V(\tilde x,\tilde y)$,
\begin{align}\label{ZV}
Z(\tilde x,\tilde y)
=\sum_{i=1}^{2N_b\!+\!1}\frac{(-1)^{i\!+\!1}
(\tilde x\!-\!\tilde x_i)}{2\sqrt{(\tilde x\!-\!\tilde x_i)^2+\tilde y^2}}
\ ,\qquad
V(\tilde x,\tilde y)
=\sum_{i=1}^{2N_b\!+\!1}\frac{(-1)^{i\!+\!1}}{2\sqrt{(\tilde x\!-\!\tilde x_i)^2+\tilde y^2}},
\end{align}
where $\tilde x_i$ are the location of the boundaries between the black/white regions in the droplet representations of the geometries and $N_b$ is the number of black or white regions with finite lengths. 
See~\cite{Jang:2016aug,Jang:2017gwd, Jang:2019pve} for functional forms of $G_{pq}$ and $F_{pqrs}$ and other detailed conventions. These functions are written in terms of the Legendre polynomials as follows~\cite{Kim:2016dzw},
\begin{align}\label{ZVfunc}
&Z(r,\xi)=\frac1{2}\Big[\xi+\sum_{n=1}^{\infty}{\rm C}_n\big[(n+1)P_{n+1}(\xi)-2\xi nP_{n}(\xi)+(n-1)P_{n-1}(\xi)\big]
\left(
\frac{2\pi\mu_{0}l_{\rm P}^{3}}{r}
\right)^{n}
\Big],\nn\\
&V(r,\xi)=\frac{1}{2r}\Big[1+\sum_{n=1}^{\infty}{\rm C}_nP_{n}(\xi)
\left(
\frac{2\pi\mu_{0}l_{\rm P}^{3}}{r}
\right)^{n}\Big],
\end{align}
where $\xi =\frac{\tilde x}r$ with 
$r = \sqrt{\tilde x^2+\tilde y^2}$, $P_n(\xi)$ are the Legendre polynomials, and we have 
introduced~\cite{Jang:2019pve}
\begin{align}\label{Cn}
{\rm C}_n = \sum_{i=1}^{2 N_b +1}(-1)^{i+1}\left(
\frac{{\tilde x_i}}{2\pi\mu_{0}l_{\rm P}^{3}}\right)^n
\end{align}
with Planck length $l_{{\rm P}}$. In order to obtain the HEE for the mABJM theory in the small mass limit, we consider the asymptotic expansion of the LLM geometries up to quadratic order in $\mu_0$.\footnote{The HEEs for the massive ABJM theory was investigated in various 
contexts~\cite{Kim:2014yca,Kim:2016dzw,Jang:2017gwd, Balasubramanian:2018qqx, Ahn:2019pqy}.} One can see in this small mass limit that physical quantities, such as vevs of CPOs with $\Delta = 1,2$ and the HEE, are completely expressed by the following two quantities  
\begin{align}\label{A2A3}
A_{2}=\frac12\left({\rm C}_2-{\rm C}_1^2\right),\qquad A_{3}=\frac13 \left({\rm C}_3-3{\rm C}_1{\rm C}_2+2{\rm C}_1^3\right).
\end{align}

Before we write the non-linear KK reduction of the 11-dimensional supergravity, we notice that the $\mathbb{Z}_k$-orbifold has no non-trivial role in the KK reduction. The reason is the following. The LLM geometry has SO(2,1)
$\times{\rm SO}(4)/\mathbb{Z}_k\times{\rm SO}(4)/\mathbb{Z}_k$ isometry and becomes asymptotically AdS$_4\times S^7/\mathbb{Z}_k$. In order to reflect such symmetry of the LLM geometry, one needs to write the asymptotic metric $S^7/\mathbb{Z}_k$ as \begin{align}\label{obfd}
ds_{S^7/\mathbb{Z}_k}^2 = d\tau^2 + \frac{ d\theta^2 + \sin^2\theta d\phi^2 + (d\psi + \cos\theta d\phi)^2}{4}+ \frac{ d\tilde\theta^2 + \sin^2\tilde\theta d\tilde\phi^2 + (d\tilde\psi + \cos\tilde\theta d\tilde\phi)^2}{4},
\end{align} 
where $(\theta,\phi,\psi)$ and 
$(\tilde\theta,\tilde\phi,\tilde\psi)$ are Euler angles with ranges, $0\le \theta,\tilde\theta\le\pi$, $0\le \phi,\tilde\phi\le 2\pi$, and 
$0\le \psi,\tilde\psi\le \frac{4\pi}{k}$ 
~\cite{Auzzi:2009es,Cheon:2011gv}. We see that in \eqref{obfd}, the two $S^1$ circles with angles $\psi$ and $\tilde\psi$ are orbifolded. However, in the asymptotic expansion 
of the LLM geometries, components of the metric and the 4-form field strength have no dependence of $\psi$ and $\tilde\psi$. For this reason,  one can follow the method of non-linear KK reduction developed in \cite{Jang:2016aug,Jang:2017gwd, Jang:2019pve}, even for the cases of $k>1$, since the presence of discrete torsion~\cite{Jang:2019pve} is originated from the $\mathbb{Z}_k$-orbifold and is only related to the coordinates $\psi$ and $\tilde \psi$.

Keeping in mind the comments in the previous paragraph, one can implement the non-linear KK reduction up to quadratic order in $\mu_0$ for the LLM geometry to obtain 
an asymptotically AdS$_4$ geometry. Then we obtain   
\begin{align}\label{AdS4-met}
ds^2 = \frac{L^2}{z^2}\left[f(z)\eta_{ij} dx^i dx^j + g(z)dz^2 \right],
\end{align}
where  $\eta_{ij} = {\rm diag}(-1,1,1)$ and 
\begin{align}
&f(z)=1- \frac1{45}\left(30 + \beta_3^2\right)(\mu_0z)^2+{\cal O}( z^4),
\nn\\
&g(z)=1-  \frac1{360}\left(960 + 29\beta_3^2\right)(\mu_0z)^2+{\cal O}( z^4).
\end{align}
Here, the quantities $\beta_3$ and the radius of the AdS$_4$ $L$ are written in terms of $A_2$ and $A_3$ as 
\begin{align}\label{Lbeta3}
\beta_3 = \frac {3A_3}{A_2^{3/2}}, \qquad
L=\frac12(32\pi^2A_2)^{1/6}l_{\mathrm{P}}.
\end{align} 
 In order to write the metric in the FG coordinate, we introduce the coordinate transformation $z\to z + \frac{\mu_0^2 }{1440}\left(960 + 29 \beta_3^2\right)  z^3$. The result is
\begin{align}\label{newFG}
ds^2 = \frac{L^2}{z^2}\left[ d z^2 + \left(1- \Big(2 +\frac{\beta_3^2}{16}\Big)(\mu_0  z)^2 + {\cal O}\Big((\mu_0 z)^4\Big)\right)\eta_{ij}dx^i dx^j\right].
\end{align}
In the next subsection, we use the asymptotically AdS$_4$ metric in  \eqref{newFG} to construct the renormalized HEE and HSC following the methods developed in the previous sections.

\subsection{Renormalized HEE and HSC in the mABJM theory}

In \cite{Jang:2017gwd}, we have shown that the metric in \eqref{newFG} is a solution to equations of motion derived from the action of Einstein gravity with negative cosmological constant coupled to two scalar fields $T$ and $\Psi$.  The action was obtained from the KK reduction of the 11-dimensional gravity on LLM background, and it is given by 
\begin{align}\label{4dact}
S = \frac{1}{16\pi G_4}\int d^4 x \sqrt{-g}\left({\cal R} - 2\Lambda\right) + S_m, 
\end{align}
where $\Lambda=-\frac{3}{L^2}$ is the negative cosmological constant and the matter action is given by
\begin{align}\label{S_m}
S_m=&-\frac{1}{32\pi G_4}\int d^4x\sqrt{-g}\Big(\nabla_p T\nabla^p T+M_t^2T^{2}+\nabla_p \Psi\nabla^p \Psi+M_\psi^2\Psi^{2}\Big).
\end{align} 
The field $\Psi$ is a genuine scalar, which is dual to the CPO of conformal dimension $\Delta=1$, with mass $M_\psi^2=\frac{\Delta(\Delta-3)}{L^2}=-\frac2{L^2}$, whereas $T$ is a pseudoscalar and it is dual to a gauge invariant operator of conformal dimension $\Delta=2$, hence has the same mass as that of $\Psi$. 

The solutions to the equations of motion of those scalar fields as well were obtained from the KK reduction of the 11-dimensional LLM solutions: 
\begin{align}\label{PsiT}
T(z)=4\mu_0z+s_1z^3+\cdots,\qquad \Psi(z)=-\frac1{\sqrt{2}}\beta_3\mu_0 z+v_1z^3+\cdots.
\end{align}
Using the conformal dimensions assignments of the previous paragraph and comparing this solution  with the general solutions we wrote in section \ref{D4}, we notice that for the field $T$ the solutions belong to the first type of solution in \eqref{gh(z)} with $s_0=4\mu_0$, whereas for the field $\Psi$ they belong to the second type of solutions in \eqref{gh(z)} with $v_0=-\frac1{\sqrt{2}}\beta_3\mu_0$.  One can also read the solution to the warp factor $h(z)$ by using \eqref{gh(z)} and \eqref{PsiT}. The answer is
\begin{align}\label{hLLM}
h(z)=-\frac{s_0^2}8z^2+\cdots-\frac{v_0^2}8z^2+\cdots=-\left(2+\frac{\beta_3^2}{16}\right)\mu_0^2z^2+\cdots.
\end{align}
As expected, this is consistent with the value of the warp factor that can be read from \eqref{newFG}. 

The parametrization in \eqref{wofzx}, which describe the embedding of the minimal surface $\Sigma_A$ into the bulk space, is convenient to separate the divergent terms from the regular terms and then propose the appropriate counter terms to cancel those divergences. For the case at hand, where the subspace $A$ is a disc of radius $R $, however, to calculate the finite value of the regularized HEE and HSC, one needs to find the re-summation of the series in \eqref{asyExp} at each order in the mass parameter $\mu_0$, which is very difficult. Therefore, we introduce   an an alternative parametrization for the embedding  as 
\begin{align}
x^p=\Big(t=t_0,z,x=\rho(z)\cos\theta, y=\rho(z) \sin\theta\Big).
\end{align} 
Though we choose the mapping \eqref{ind-met}, which is different from that of \eqref{wofzx}, the resulting counter terms for the HEE and the HSC have the same forms since they are independent of coordinate choices.

Denoting the coordinates on the minimal surface as $\sigma^\alpha=(\theta,\rho)$, the induced metric $\gamma_{\alpha\beta}=\partial_\alpha x^p\partial_\beta x^q g_{pq}$ becomes
\begin{align}\label{ind-met}
\gamma_{\rho\rho}=\frac{L^2}{z^2}\left[\frac1{\rho'(z)^2} +\big(1+h(z)\big)\right], \qquad\gamma_{\theta\theta} =\frac{L^2}{z^2}\left(1+h(z)\right) \rho(z)^2,
\end{align}
where we have used the bulk metric $g_{pq}$ in \eqref{newFG} and
 the warp factor is read from \eqref{hLLM}
\begin{align}
1+h(z)=1+h_0z^\alpha+h_1z^{2\alpha}+\cdots,\quad {\rm with}~\alpha=2,~h_0=-\left(2+\frac{\beta_3^2}{16}\right)\mu_0^2.
\end{align}
The area of the minimal surface is given by  
\begin{align}\label{ASigma}
{\cal A}_{\Sigma_A}
=\int d\rho\int d\theta\sqrt{\det\gamma_{\alpha\beta}}
={2\pi L^{2}}\int_{0}^{z_m}d z{\cal L}_{A},
\end{align}
where 
\begin{align}
{\cal L}_{A}=\frac{\rho(z)}{z^2}\sqrt{1+h(z)+\Big(\big(1+h(z)\big) \rho'(z)\Big)^2}.
\end{align}
In order to determine $\rho(z)$ which  minimize the area ${\cal A}_{\Sigma}$, we solve the  Euler-Lagrangian equation
for ${\cal L}_{A}$, 
order by order in $\mu_0$ with the boundary condition $\rho(z=0)=R $. Up to quadratic order in $\mu_0$, the result is~\cite{Kim:2014yca, Kim:2016dzw, Jang:2017gwd} 
\begin{align}
\rho(z)=\rho_0(z)+\rho_2(z)\mu_0^2+{\cal O}(\mu_0^4),
\end{align}
where
\begin{align}\label{hofz}
&\rho_0(z)=\sqrt{R ^2-z^2},\nn\\
&\rho_2(z)=\frac{\mu_0^2}{6\sqrt{R ^2-z^2}}\left(4+\frac{\beta_3^2}{8}\right)
\left[-\frac{z^4}2+
2R ^2z^2-4R ^3z+4R ^4\log\Big(\frac{R +z}{R }\Big)
\right].
\end{align}
The value of the turning point $z_m$ is determined by $\rho'(z_{m})\to\infty$ and is given by
\begin{align}
z_{m}
=
R 
-\frac{R ^{3}}{6}
\left(2+\frac{\beta_3^2}{16}\right)\Big(
5
-8\log2
\Big)
\mu_{0}^{2}
+{\cal O}(\mu_0^4).
\end{align}

Using these results in \eqref{ASigma} and introducing a cut-off $z=\epsilon$, we can calculate the HEE up to quadratic in $\mu_0$
\begin{align}\label{S_A2}
S_{A}^{{\rm reg}}
= &
\frac{2\pi L^{2}}{4 G_4}\int_{\epsilon}^{z_m}d z\,{\cal L}_A=
\frac{\pi L^{2}}{2 G_4}
\left(
\frac{R }{\epsilon}
-1
-\frac{32+\beta_{3}^{2}}{24}R ^{2}\mu_{0}^{2}
+\cdots\right),
\end{align}
where here and in the following equations, the ellipses denote terms which are higher order in $\mu_0$.
Since we are considering the $\alpha=2$ case, only the first counter term in \eqref{CT1-4} is required to cancel the divergences. The induced metric on the boundary curve $\partial\Sigma_A^{\epsilon}$ is 
\begin{align}\label{gammatt}
\tilde\gamma_{\theta\theta}={L^2}\frac{1+h(\epsilon)}{\epsilon^2}\rho(\epsilon)^2.
\end{align}
Then counter term is
\begin{align}
S^{(1)}_{\rm ct}=-\frac{L}{4G_4}\int d\theta\sqrt{\tilde\gamma_{\theta\theta}}=
-\frac{\pi L^{2}}{2 G_4}
\frac{R }{\epsilon}+{\cal O}(\epsilon^\alpha).
\end{align}
From the renormalized HEE for $\alpha>1$, which was constructed in \eqref{Area4d}, we obtain
\begin{align}\label{S_A2-1}
S^{\rm ren}_{A}=
-\frac{\pi L^{2}}{2 G_4}
\left(1+\frac{32+\beta_{3}^{2}}{24}R ^{2}\mu_{0}^{2}
+\cdots\right).
\end{align}
Here the negative sign of the contribution from the mass deformation (a relevant perturbation) is related to the $F$-theorem in the 3-dimensional CFT.

Similarly, the HSC, which is identified with the volume ${\cal B}_\epsilon$ enclosed by the disc $A$ and the static minimal area surface $\Sigma_A^{\epsilon}$ at  a constant time slice $t=t_0$, is given by
\begin{align}
{\cal C}_{A}^{{\rm reg}}
=\frac{V({\cal B}_\epsilon)}{8\pi LG_4}.
\end{align}
See Fig.1. Using the bulk coordinates $x^m=(z,\rho,\theta)$ on the constant time slice, the volume is given by
\begin{align}\label{volume-gamma}
{\cal C}_{A}^{{\rm reg}}
=
&\frac{L^2}{8\pi G_4}\int_{0}^{2\pi}d\theta\int_{\epsilon}^{z_m}dz\int_{0}^{\rho(z)}d\rho \frac{\rho}{z^{3}}\big[1+h(z)\big]
\nn\\
&=
\frac{\pi L^{2}}{16\pi G_4}
\left[
-1
+\frac{R ^{2}}{\epsilon^{2}}
+2\log
\left(
\frac{\epsilon}{R }
\right)
+\left(\frac{32+\beta_3^2}{8}\right)
\left(
1+\log
\left(
\frac{\epsilon}{R }
\right)
\right)
R ^{2}\mu_{0}^{2}
+\cdots\right].
\end{align}

The above result contains three divergent terms.  Those divergences are regulated by the three counter terms that were introduced in the subsection \ref{RHSC4}. The first counter term is given by
\begin{align}
{\cal C}_{\rm ct}^{(1)}&=-\frac{1}{16\pi G_4}\int_0^{2\pi}d\theta \int_{0}^{\rho(\epsilon)}d\rho\sqrt{\det h_{ij}}=-\frac{ L^2}{8 G_4}\frac{1+h(\epsilon)}{\epsilon^2}\int_{0}^{\rho(\epsilon)}\rho d\rho\nn\\
&=\frac{L^{2}}{16 G_4}\left[-\frac{R ^2}{\epsilon^2}+1+\left(\frac{32+\beta_3^2}{16}\right)R ^2\mu_0^3+\cdots\right].
\end{align}
Recalling that in the case we are considering here, $\alpha=2$, the second counter term is given by
\begin{align}
{\cal C}_{\rm ct}^{(2)}&= \frac{1}{8\pi G_4}\ln\Big(\frac\epsilon R \Big)\int_0^{2\pi} d\theta \int_{0}^{\rho(\epsilon)}d\rho\sqrt{\det h_{ij}}\Big(1+\frac L2 K\Big)=\frac{\pi L^2}{8\pi G_4}\ln\Big(\frac\epsilon R \Big)\int_{0}^{\rho(\epsilon)}d\rho \frac{\rho}{\epsilon}h'(\epsilon)\nn\\
&=-\frac{ L^2}{16 G_4}\left[\ln\Big(\frac\epsilon R \Big) \left(\frac{32+\beta_3^2}8\right) R ^2\mu_0^2+\cdots\right]. 
\end{align}
In order to calculate the third counter term, we need to obtain the extrinsic curvature of the boundary curve $\partial\Sigma_A^{\epsilon}$ using the embedding $x^i=(\rho=\rho(\epsilon),\theta)$. The induced metric and the extrinsic curvature are given by
\begin{align}
&\tilde\gamma_{\theta\theta}={L^2}\frac{1+h(\epsilon)}{\epsilon^2}\rho(\epsilon)^2,\qquad \tilde K_{\rho\rho}=0,\quad\tilde K_{\rho\theta}=\tilde K_{\theta\rho}=0,\nn\\
&\tilde K_{\theta\theta}=-\frac{L\rho(\epsilon)}{\epsilon}\sqrt{1+h(\epsilon)},\qquad \tilde K=g^{ij}\tilde K_{ij}=-\left(\frac{L\rho(\epsilon)}{\epsilon} \sqrt{1+h(\epsilon)}\right)^{-1}.
\end{align}
Therefore, the third counter term is
obtained as\begin{align}
\tilde{\cal C}_{\rm ct}&=\frac{L^2}{16\pi G_4}\ln\Big(\frac\epsilon R \Big)\int_0^{2\pi}d\theta\sqrt{\det\tilde\gamma_{\theta\theta}}\tilde K=-\frac{ L^2}{8G_4}\ln\Big(\frac\epsilon R \Big).
\end{align}
Note that, unlike the the first two counter terms, this one is exact and has no $\mu_0$ corrections, which means that it is independent of the scalar deformation.   This is because, when we deform the pure AdS$_4$ space to obtain an asymptotically AdS$_4$ space by adding scalar deformations, the induced metric on the cut-off surface $\partial {\cal M}_\epsilon$  as well as those on the minimal surface $\Sigma_A$ get deformed, however, the boundary curve $\partial\Sigma_A$ of the minimal surface remains the same. Since the third counter term is built by the invariants on this codimension-three boundary curve, it is independent of the scalar deformation. In general, as we have shown by using the examples of $d=3$ and $d=4$ in the previous section, there is always one divergent term which is independent of the scalar deformation, it was logarithmic in $d=3$ and ${\cal O}(\epsilon^{-1})$ in $d=4$. The counter term which cancels this divergence is always built from the invariants on the codimension-three  boundary of the minimal hyper-surface $\Sigma_A$. Therefore, the general claim in the literature, which states that the counter terms to cancel all the divergences encountered in the complexity calculations are built from the invariants on the codimension-two cut-off hyper-surface $\partial {\cal M}_\epsilon$, does not account for this particular divergence.

Finally, adding the three counter terms to the regularized HSC in \eqref{volume-gamma} using the renormalized HSC constructed in \eqref{HSC7} and \eqref{ctct4}, we  obtain 
\begin{align}\label{volume-gamma-1}
{\cal C}^{\rm ren}_{A}
=
\frac{ 3L^{2}}{16^2 G_4}
\left(
 (32+\beta_3^2)R ^{2}\mu_{0}^{2}
+\cdots\right),
\end{align}
where we notice that the renormalized HSC is vanishing in the absence of the relevant perturbation. Therefore, the contribution for the HSC by the mass deformation (a relevant deformation) is always positive, while that for the HEE is negative.  From the renormalized HEE in \eqref{S_A2-1}, we obtain the relation for the disk entangling region up to $\mu_0^2$-order, 
\begin{align}\label{dcds}
\Delta {\cal C}_A^{{\rm ren}} = -\frac{9}{16\pi^2} \Delta S_A^{{\rm ren}}, 
\end{align}
where  $ \Delta S_A^{{\rm ren}}$ and $\Delta {\cal C}_A^{{\rm ren}}$, respectively, are the variations of the renormalized HEE and HSC due to the relevant perturbations. The properties of the relation \eqref{dcds} were investigated in \cite{Momeni:2017ibg}. 
However, in our case, it is not clear that the relation \eqref{dcds} is satisfied for higher $\mu_0$-orders and different shapes of subregions.

\section{Conclusion} 
In this paper, we examined the renormalization of HEE and HSC of general entangling subregions on the asymptotically AdS$_4$ and AdS$_5$ geometries under relevant perturbations originated from a bulk scalar field. We considered the HSC of the CV conjecture and omitted the case of the asymptotically AdS$_3$ geometry, which is similar to the case of the AdS$_5$ geometry.  In order to renormalize these quantities in a coordinate independent way, we explicitly constructed universal counter terms using the  holographic renormalization method. 

 For the divergences of the HEE on an asymptotically AdS$_{d+1}$ geometry, the proposed counter terms are integrals of the curvature invariants on the $(d-2)$-dimensional boundary of the RT minimal hyper-surface. We pointed out that curvature invariants on the boundary of the RT minimal hyper-surface are independent of the bulk stress tensor. On the other hand, the HEE contains subleading divergences whose coefficients  are determined by the back reaction of the stress tensor on the geometry. We showed that the counter terms that cancel  these subleading divergence,  must contain invariants of the bulk matter fields in addition to the curvature invariants on the boundary of the RT hyper-surface. We have determined the exact forms of these counter terms in the asymptotically AdS$_4$ and AdS$_5$ geometries with arbitrary shapes of entangling regions.
 
 Taking lesson from the renormalization of the HEE, the counter terms for the divergences in HSC were proposed as integrals of the curvature invariants on the $(d-1)$-dimensional cut-off hyper-surface at $z=\epsilon$, with $z$ being the holographic coordinate.  
 In this case the curvature invariants on the cut-off hyper-surface are dependent on the bulk stress tensor. Therefore, it looks natural to build the counter terms just from the integrals of the curvature invariants on the  $(d-1)$-dimensional cut-off boundary.  However, we pointed out that there is always a logarithmic divergence for odd $d$ and a power-law divergence for even $d$, which are expressed in terms of integrals of curvature invariants on the boundary of the RT minimal hyper-surface. We argued that the existence of this special divergence is attributed to the fact that the $(d-1)$-dimensional cut-off boundary
meets the $(d-2)$-dimensional boundary of the RT hyper-surface and gets the UV divergence. 
 We showed that the complete counter terms for the divergences of HSC are expressed as integrals of the curvature invariants on the $(d-1)$-dimensional cut-off boundary plus integrals of the curvature invariants on the $(d-2)$-dimensional boundary of the RT hyper-surface.

We have tested our general construction of the renormalized HEE and HSC for an asymptotically AdS$_4$ geometry, which was obtained from the non-linear KK reduction of the 11-dimensional LLM geometry. We obtained  coordinate independent finite results for both HEE and HSC with a disk shape of entangling region.
For our convenience of the coordinate choice for the disk, we used a different mapping with \eqref{wofzx}. However, the counter terms for the HEE and HSC have the same form with those in subsections \ref{D4} and \ref{RHSC4}, respectively. 

Intriguingly, we found that the coordinate independent renormalization of the HSC in the asymptotically AdS$_5$ is not possible in the range $0<\alpha<1$, i.e., $0< \Delta < \frac{1}{2}\,\, {\rm and} \,\, \frac{7}{2}< \Delta < 4$ of the relevant operators in the 4-dimensional dual field theory. That is,  a divergent term in that range of $\alpha$ cannot be cancelled out by adding any curvature invariant.   We also  noticed that the  case  of $\frac{7}{2}< \Delta < 4$ does not violate the unitary bound $(\Delta \ge1)$ for primary operators. Therefore, the problem of the non-renormalizability of the HSC in the asymptotically AdS$_5$ is genuine in this range of $\alpha$.   It will be interesting if one figures out the physical reason of this phenomenon by investigating cases of other dimensions  and other HSC conjectures, for instance,  the HSC in the CA conjecture. 

\section*{Acknowledgments}

 We would like to thank Kyung Kiu Kim and  Chanyong Park for useful comments and discussions. This work was supported by the National Research Foundation of Korea(NRF) grant with grant number NRF-2018R1D1A1B07048061 (D.J.),     NRF-2019R1F1A1056815 (Y.K.),  NRF-2017R1D1A1A09000951, NRF-2019R1F1A1059220, NRF-2019R1A6A1A10073079 (O.K.), and NRF-2017R1D1A1B03032523 (D.T.).
 Y.K. and O.K. were also supported by NRF Bilateral Joint Research Projects (NRF-JSPS collaboration)
``String Axion Cosmology''.”

\end{document}